\magnification=1200
\baselineskip=18truept

\line{\hfill RU-98-02}

\vskip 2truecm
\centerline{\bf Remark on lattice BRST invariance.}

\vskip 1truecm
\centerline{Herbert Neuberger}
\vskip .5truecm

\centerline {Department of Physics and Astronomy}
\centerline {Rutgers University, Piscataway, NJ 08855-0849}

\vskip 1.5truecm

\centerline{\bf Abstract}
\vskip 0.75truecm
A recently claimed resolution to the lattice Gribov problem in the context
of chiral lattice gauge theories is examined. Unfortunately, I find that
the old problem remains. 

\vfill\eject

Within the BRST framework, perturbation theory indicates that the trivial
Gaussian fixed point associated with an asymptotically free chiral
four dimensional gauge theory has only one (marginally) unstable direction
associated with the gauge coupling. If the requirement of BRST 
invariance is relaxed, one has a larger, but finite, number of
unstable directions, and the approach to the desired continuum limit
is conditional on 
fine tuning to eliminate the effects of the extra unstable directions.
The fine tuning can be specified imposing additional 
identities, one for each extra unstable direction. One chooses
identities that would hold if BRST invariance were exact. The
RG then indicates that full BRST invariance will be restored in the
continuum limit, and that a self-consistent, potentially strongly coupled,
chiral gauge theory exists also outside perturbation theory. This
is my understanding of the essence of the ``Rome proposal''[1]. About a decade ago, when 
lattice chiral 
gauge theory still amounted mostly
to an industry of failures, the Rome 
proposal triggered a process of restoring
common sense to this subfield. 

The one nagging suspicion about the Rome proposal that I have is the lack
of certainty surrounding BRST invariance at the non-perturbative level
in simpler, non-chiral, lattice gauge theory. 
This, in addition to some lack
of clarity regarding global anomalies [2], indicates that even in principle,
the Rome proposal may need a (hopefully small) upgrade or ``bug-fix'' [3]. 
In the rest of this note I shall have nothing more to say about global
anomalies and shall focus only on non-perturbative BRST invariance.

In the late seventies to mid eighties, after some confusion having to do with
Gribov copies, it became clear that as long as all solutions to the
gauge fixing constraints were included together with the associated full
Faddeev-Popov determinants (with their signs) no formal problems of principle
were evident [4]. Unfortunately, using a direct transplant to the lattice [5]
of the  continuum BRST operation (interpreted in the language of Lie Algebra 
Cohomology with the addition of 
a Koszul-Tate resolution), it was found that all 
BRST invariant observables have the form ${0\over 0}$ [6]. This would imply,
in particular, that applying the Rome approach to QCD for example, 
produces an ill defined theory. 
Moreover, one cannot imagine a lattice RG
transformation that produces a fixed point
and a one dimensional renormalization group
trajectory
emanating from it in the one marginal
direction available, along which all actions are
BRST invariant, corresponding to an anomaly
free, asymptotically free, chiral gauge theory. 
Note that the BRST transformation is always a local
operator, since it simply encodes local gauge invariance.
While these observations do
not constitute proof that the Rome proposal has a defect, it seems to me that they do provide 
grounds to ask for an unequivocal resolution. 

In a stream of recent publications [7] it has been repeatedly 
claimed that the above problem 
can be avoided by choosing a special 
gauge fixing term in the Lagrangian. 
The term depends only on the
gauge fields, has an absolute minimum when all lattice parallel
transporters equal unity, and is naturally written as the sum of two terms
rather than the square of a single local factor. Since most 
of this work is directed to the $U(1)$ case, the issue of asymptotic freedom
being somewhat divorced from the problems encountered with
lattice chirality, I shall restrict myself to the $U(1)$ case below.

As far as I can see, the gauge fixing employed in [7] fits into
the lattice BRST framework of [5,6] exactly
as well as the original proposal [1]. 
The somewhat different structure of the gauge fixing term
is irrelevant to the lattice BRST problem I am addressing.
For example, even with the gauge fixing
term of [7], BRST invariant 
lattice QED will still produce ${0\over 0}$ for 
the expectation value of any BRST
invariant observable.  
The unfortunate 
conclusion is that no progress of principle 
beyond the Rome proposal has
been made in [7]. I am leaving open 
the possibility that I am making a mistake
(for which I wish to apologize in advance) 
because the argument below is so simple that it is hard to imagine it
being overlooked in all the papers in [7].

The basic issue boils down to this: Consider the full
action with all Grassmann variables set to zero. 
This action splits additively into a gauge invariant 
term and a gauge fixing term. The question now is
whether it is possible, by the addition of terms 
involving only BRST ghosts, to restore full BRST
invariance. The answer is positive both in the case
of [1] and in the case of [7]. When the non-ghost
fermion fields are turned on BRST invariance is
lost both in [1] and in [7]; but this is besides 
the point I am making here. The easiest way to
check a particular case
is to go through the {\it formal} steps one would take 
to implement the Faddeev Popov trick
on the lattice.
While the steps contain an error when viewed
as identities they do produce 
a path integral that could
be taken as the {\it definition of
a BRST invariant, nonperturbatively defined, 
field theory}. As explained
in [6], if the steps were
mathematically correct we would, of course,
end up with a logical contradiction to the result of [6],
since a perfectly reasonable lattice gauge invariant
theory would be proven equivalent to a nonsensical one.
In the next few lines I sketch these formal
steps for the gauge fixing term of [7]. The conclusion
would be that the gauge fixing term of [7] 
is not special
at all, and can be incorporated in standard
lattice BRST, as used in [1] and in [5,6].

The gauge fixing term of [7], appearing with a minus sign in the
exponent in the standard formula for the partition function ($\alpha >0 $), 
can be written as
$$
{1\over {2\alpha}}\sum_x S(U;x ),\eqno{(1)}$$
where $x$ denotes lattice sites, and $U$ the collection of $U(1)$ link
variables $U_\mu (y)$.  $S(U,x)$ is local 
in that it depends only on $U_\mu (y)$ with $y$ near $x$. For 
all $U$, $S(U,x) \ge 0$ and $S({\bf 1},x ) = 0$ for all $x$. 
Since the  $U_\mu (y)$ are compact there exists a number $M > 1$, such
that for all $U$ and $x$,  $S(U,x) \le M-1$. Let $f(U,x)$ be a gauge fixing
function defined as follows:
$$
f(U,x) =\sqrt {M+S(U,x)}.\eqno{(2)}$$
In (2) I chose the positive branch of the square root. 
Clearly, $f(U,x)$ is a uniformly converging series in $S(U,x)$. Introducing
a product of delta functions,  $\prod_x \delta (b (x) - f(U,x))$,
into the path integral of a gauge invariant model, adding the Faddeev-Popov 
determinant factor, and averaging over $b(x)$ with a Gaussian 
weight $\exp {[-{1\over{2\alpha}} \sum_x b^2 (x)]}$, produces the bosonic
gauge fixing part of equation 1 up to an irrelevant multiplicative 
constant $\exp {[-{1\over{2\alpha}} \sum_x M]}$.
The result of [6] directly applies and leads to the ${0\over 0}$ problem.
For this case the ghosts enter only bilinearly
so the alternative explanation offered in [6] also applies:
the number of solutions to the gauge fixing equations is 
even (including 0) {\it generically}.
On the trivial orbit there is a single solution because it is non-generic,
being a minimum. 
However, almost on all orbits in the vicinity of the trivial orbit
there will be several solutions making contributions that sum up to zero.

\bigskip
\centerline{\bf  Acknowledgments.}
\medskip

This research was supported in part by the DOE under grant \#
DE-FG05-96ER40559. I am indebted to M. Testa 
who anticipated the result of this note 
in a question he posed to me during the last lattice
conference; unfortunately I was too dense to see immediately then
that he was right. I also wish to thank Y. Kikukawa for a 
discussion. In addition I wish to thank
J. Smit for alerting me to possible
misunderstandings that a somewhat more tersely
worded earlier version of this note could 
have triggered. 


\bigskip
\centerline{\bf  References.}
\medskip

\item{[1]} A. Borelli, L. Maiani, G.-C. Rossi, R. Sisto and M. Testa,
Nucl. Phys. B333 (1990) 335; L. Maiani, G.-C. Rossi and M. Testa,
Phys. Lett. 292B (1992) 397.
\item{[2]} E. Witten, Phys. Lett. B 117 (1982) 324.
\item{[3]} H. Neuberger, Nucl. Phys. B. (Proc. Suppl.) 17 (1990) 17.
\item{[4]} P. Hirschfeld, Nucl. Phys. B 157 (1979) 37; B. Sharpe,
J. Math Phys. 25 (1984) 3324.
\item{[5]} H. Neuberger, Phys. Lett. B 175 (1986) 69.
\item{[6]} H. Neuberger, Phys. Lett. B 183 (1987) 337.
\item{[7]} Y. Shamir, Phys. Rev. D57 (1998) 132; M. Golterman and Y. Shamir,
Phys. Lett. 399B (1997) 148; M. Bock, M. Golterman and Y. Shamir,
hep-lat/9709154, hep-lat/9708019, hep-lat/9709113, hep-lat/9709115,
hep-lat/9801018.

\vfill\eject\end